\documentclass[twocolumn,showpacs,preprintnumbers,amsmath,amssymb]{revtex4}

\usepackage{graphicx}
\usepackage{dcolumn}
\usepackage{bm}

\begin{document}

\preprint{APS/xxx}

\title{Scaling Invariance in Spectra of Complex Networks: A Diffusion
Factorial Moment Approach\\}

\author{Fangcui Zhao}
\affiliation{College of Life Science and Bioengineering, Beijing University of Technology, Beijing 100022, China\\}%

\author{Huijie Yang}
\altaffiliation {Corresponding author}
\email{huijieyangn@eyou.com}
\author{Binghong Wang}%
\affiliation{%
Department of Modern Physics and Nonlinear Science Center,
University of Science and Technology
of China, Hefei Anhui 230026, China\\}%

\date{\today}

\begin{abstract}
 A new method called diffusion factorial moment (DFM) is
used to obtain scaling features embedded in spectra of complex
networks. For an Erdos-Renyi network with connecting probability
$p_{ER} < \frac{1}{N}$, the scaling parameter is $\delta = 0.51$,
while for $p_{ER} \ge \frac{1}{N}$ the scaling parameter deviates
from it significantly. For WS small-world networks, in the special
region $p_r \in [0.05,0.2]$, typical scale invariance is found.
For GRN networks, in the range of $\theta\in[0.33,049]$, we have
$\delta=0.6\pm 0.1$. And the value of $\delta$ oscillates around
$\delta=0.6$ abruptly. In the range of $\theta\in[0.54,1]$, we
have basically $\delta>0.7$. Scale invariance is one of the common
features of the three kinds of networks, which can be employed as
a global measurement of complex networks in a unified way.
\end{abstract}

\maketitle

\section{\label{sec:level1}Introduction}
In recent years, complex networks attract special attentions from
diverse fields of research [1]. Though several novel measurements,
such as degree distribution, shortest connecting paths and
clustering coefficients, have been used to characterize complex
networks, we are still far from complete understanding of all
peculiarities of their topological structures. Finding new
characteristics is still an essential role at present time.

Describe the structure of a complex network with the associated
adjacency matrix. Map this complex network with $N$ nodes to a
large molecule, the nodes as atoms and the edges as couplings
between the atoms. Denote the states and the corresponding site
energies of the atoms with $\left\{ {\left| 1 \right\rangle
,\left| 2 \right\rangle ,...\left| N \right\rangle } \right\}$ and
$\{\varepsilon _1 ,\varepsilon _2 ,\ldots \varepsilon _N \}$,
repsectively. Consider a simple condition where the Hamiltonian of
the molecule reads,

\begin{equation}
 H_{mn} = (H_0 )_{mn} + (H_{coupling} )_{mn} ,
\end{equation}

\noindent where,

\begin{equation}
\begin{array}{l}
 (H_0 )_{mn} = \varepsilon _0 \delta _{mn} , \\
 (H_{coupling} )_{mn} = \left\{ {{\begin{array}{*{20}c}
 {1 - \delta _{mn} } \hfill & {connected} \hfill \\
 0 \hfill & {disconnected} \hfill \\
\end{array} }} \right. \\
 \end{array}
\end{equation}
By this way a complex network is mapped to a quantum system, and
the corresponding associated adjacency matrix to the Hamiltonian
of this quantum system.

The structure of a complex network determines its spectrum. The
characteristics of this spectrum can reveal the structure
symmetries, which can be employed as global measurements of the
corresponding complex network [2-12]. In our recent papers
[13-16], several temporal series analysis methods are used to
extract characteristic features embedded in spectra of complex
networks.

In the present paper, a new concept, called diffusion factorial
moment (DFM), is proposed to obtain scale features in spectra of
complex networks. It is found that these spectra display scale
invariance, which can be employed as a global measurement of
complex networks in a unified way. It may also be helpful for us
to construct a unified model of complex networks.

\section{\label{sec:level1}Diffusion Factorial Moment (DFM)}
Represent a complex network with its adjacency matrix: $A(G)$. The
main algebraic tool that we will use for the analysis of complex
networks will be the spectrum, i.e., the set of eigenvalues of the
complex network's adjacency matrix, called the spectrum of the
complex network, denoted with $\left\{ {E_m \left| {m = 1,N}
\right.} \right\}$. Connecting the beginning and the end of this
spectrum, we can obtain a set of delay register vectors as [13],

\[
\left\{ {E_1 - E_0 ,E_2 - E_1 ,\ldots E_n - E_{n - 1} } \right\}
\]

\[
\left\{ {E_2 - E_1 ,E_3 - E_2 ,\ldots E_{n + 1} - E_n } \right\}
\]

\[
 \vdots
\]

\[
 \left\{ {E_{N} - E_{N -1}}
,{E_0 - E_{N}} , \cdots {E_{n - 2} - E_{n - 3}} \right\}
\]

\begin{equation}
\left\{ {E_0 - E_N ,E_1 - E_0 ,\ldots E_{n - 1} - E_{n - 2} }
\right\}
\end{equation}

Considering each vector as a trajectory of a particle during $n$
time units, all the above vectors can be regarded as a diffusion
process for a system with $N + 1$ particles [17]. Accordingly, for
each time denoted with $n$ we can reckon the distribution of the
displacements of all the particles as the state of the system at
time $n$.

  Dividing the possible range of displacements into $M_0 $ bins,
the probability distribution function (PDF) can be approximated
with $p_m \approx \frac{K_m}{\sum\limits_m {K_m } }$, where $K_m
(n)\left| {m = 1,2,...M_0 } \right.$ is the number of particles
whose displacements fall in the $m$'th bin at time $n$. To obtain
a suitable $M_0 $, the size of a bin is chosen to be a fraction of
the variance, $\varepsilon = \sqrt {\frac{\sum\nolimits_{k = 1}^{N
+ 1} {(E_k-E_{k-1} )^2} }{N + 1}} $.

  If the series constructed with the nearest neighbor level spacings,
$\left\{ {E_1 - E_0 ,E_2 - E_1 ,\ldots E_0 - E_N } \right\}$, is a
set of homogeneous random values without correlations with each
other, the PDF should tend to be a Gaussian form when the time $n$
becomes large enough. Deviations of the PDF from the Gaussian form
reflect the correlations in the time series. Here, we are
specially interested in the scale features in spectra of complex
networks.

 Generally, the scale features in
spectra of complex networks can be described with the concept of
probability moment (PM) defined as [18],

\begin{equation}
 C_q (n) = \sum\limits_{m = 1}^{M_0 } {(p_m )^q} ,
\end{equation}
where $p_m$ is the probability for a particle occurring in the
$m$'th bin. Assume the PDF takes the form,

\begin{equation}
 p_m (n) = \frac{1}{n^\delta } \cdot F\left( {\frac{m}{n^\delta
}} \right).
\end{equation}
An easy algebra leads to,

\begin{equation}
 \ln C_q (n) = A + \delta \cdot (1 - q)\ln(n)
\end{equation}

  If the considered series is completely uncorrelated, the resulting
  diffusion process will be very close to the condition of
  ordinary diffusion, where $\delta=0.5$ and the function $F\left( {\frac{m}{n^\delta
}} \right)$ in the PDF is a Gaussian function of
${\frac{m}{n^\delta }}$. $\delta \ne 0.5$ can reflect the
departure of the diffusion process from this ordinary diffusion
condition [19].  The extreme condition is the ballistic diffusions
, whose PDFs read $p_m(n)=\frac{1}{n} \cdot F\left(
\frac{m}{n}\right)$. The values of $\delta$ at this condition are
$1$.

  But the approximation of PDF, $p_m \approx
\frac{K_m}{\sum\limits_m {K_m } }$, in the above computational
procedure will induce statistical fluctuations due to the finite
number of particles, which may become a fatal problem when we deal
with the spectrum of a complex network. The dynamical information
may be merged by the strong statistical fluctuations completely.
Capturing the dynamical information from a finite number of cases
is a non-trivial task.

 This problem is firstly considered by A. Bialas and R. Peschanski in analyzing the process of
high energy collisions, where only a small number of cases can be
available. A concept called factorial moment (FM) is proposed to
find the intermittency (self-similar) structures embedded in the
PDF of states [18,20-24]. The definition of FM reads,

\begin{equation}
 F_q (M) = \sum\limits_{m = 1}^{M } {J_m\cdot
(J_m-1)\cdot\cdot\cdot(J_m-q+1)},
\end{equation}
where $M$ is the number of the bins the displacement range being
divided into and $J_m$ the number of particles whose displacements
fall in the $m$'th bin.

  Stimulated by the concept of FM, we propose in this paper a new concept called diffusion factorial moment (DFM),
which reads,

\begin{equation}
 DFM_q(n) = \sum\limits_{m=1}^{M_0} {K_m \cdot
(K_m - 1) \cdots (K_m - q + 1)} .
\end{equation}
Herein we present a simple argument for the ability of DFM to
filter out the statistical fluctuations due to finite number of
cases [20,21].The statistical fluctuations will obey Bernoulli and
Poisson distributions for a system containing uncertain and
certain total number of particles, respectively. For a system
containing uncertain total number of particles, the distribution
of particles in the bins can be expressed as,

\begin{equation}
\begin{array}{l}
 {\begin{array}{*{20}c}
 \hfill \\
\end{array} }Q\left( {K_1 ,K_2 ,\cdots ,K_{M_0 } \left| {p_1 ,p_2 \cdots
,p_{M_0 } } \right.} \right) \\
 =\frac{K!}{K_1 !K_2 !\cdots K_{M_0 } !}p_1^{K_1 } p_2^{K_2 } \cdots p_{M_0
}^{K_{M_0 } }, \\
 \end{array}
\end{equation}
where $K=K_1+K_2+\cdots+K_{M_0}$. Hence,
\begin{equation}
\begin{array}{l}
 \left\langle {K_m (K_m - 1) \cdots (K_m - q + 1)} \right\rangle \\
 = \int {dp_1 dp_2 \cdots dp_{M_0 } P(p_1 ,p_2 , \cdots ,p_{M_0 } )\cdot }
\\
 {\begin{array}{*{20}c}
 \hfill \\
\end{array} }\sum\limits_{K_1 } {\sum\limits_{K_2 } { \cdots
\sum\limits_{K_{M_0 } } Q } } \left( {K_1 ,K_2 , \cdots ,K_{M_0 }
\left| {p_1 ,p_2 ,} \right. \cdots ,p_{M_0 } } \right)\cdot\\
{\begin{array}{*{20}c}
 \hfill \\
\end{array} }
K_m(K_m - 1) \cdots(K_m - q + 1) \\
 = K(K - 1)(K - 2) \cdots (K - q + 1)\int {dp_1 dp_2 \cdots dp_{M_0 }}\cdot\\
{\begin{array}{*{20}c}
 \hfill \\
\end{array} }
P(p_1 ,p_2 , \cdots ,p_{M_0 } ) \cdot p_m^q \\
 = K(K - 1) \cdots (K - q + 1)\left\langle {p_m^q } \right\rangle \\
 \end{array}
\end{equation}
That is to say,

\begin{equation}
 DFM_q (n)\propto C_q (n)
\end{equation}
And consequently, $Eq.(6)$ becomes,

\begin{equation}
 \ln DFM_q (n) = C+\ln C_q(n)=B + \delta \cdot (1 - q)\ln(n)
\end{equation}
Therefore, DFM can reveal the strong dynamical fluctuations
embedded in a time series and filter out the statistical
fluctuations effectively. We will use the DFM instead of the PM to
obtain the scale features in spectrum of a complex network.

  It should be pointed out that the scale features in our DFM is
completely different from that in FM. The FM reveals the
self-similar structures with respect to the number of the bins the
possible range of the displacements being divided into, i.e., the
scale is the displacement. In DFM, the considered scale is the
time $n$. At time $n$, the state of the system is,
$(E_n-E_0,E_{n+1}-E_1,E_{n+2}-E_2,\cdots,E_N-E_{N-q},E_0-E_{N-q+1})$.

  In one of our recent works [13], joint use of the detrended
fluctuation approach (DFA) and the diffusion entropy (DE) is
employed to find the correlation features embedded in spectra of
complex networks. In that paper we review briefly the relation
between the scale invariance exponent, $\delta$, and the
long-range correlation exponent $\alpha$. For fractional Brownian
motions (FBM) and Levy walk processes, we have $\delta=\alpha$ and
$\delta=\frac{1}{3-2 \cdot \alpha}$, respectively. Generally, we
can not derive a relation between these two exponents. Herein, we
present the relation between the concepts of DFM and DE. From the
probability moment in $Eq.4$ we can reach the corresponding
Tsallis entropy, $S_{Tsallis}$, which reads,

\begin{equation}
 S_{Tsallis}(q) = \frac{1-\sum\limits_{m = 1}^{M_0 } {(p_m
 )^q}}{1-q}=\frac{1-C_q(n)}{1-q}.
\end{equation}
A trivial computation leads to the relation between the DE,
(denoted with $S_{DE}$), the PM and the Tsallis entropy, as
follows,

\begin{equation}
\begin{array}{l}
 S_{DE}=
  \mathop {\lim }\limits_{q \to {1} }\frac{1-\sum\limits_{m = 1}^{M_0 } {(p_m
 )^q}}{1-q} \\
 =\mathop {\lim }\limits_{q \to {1} }\frac{1-C_q(n)}{1-q}
 =\mathop {\lim }\limits_{q \to {1} } S_{Tsallis}(q).
 \end{array}
\end{equation}
Hence DFM can detect multi-fractal features in spectra of complex
networks by adjusting the value of $q$. The DE is just a special
condition of DFM with $q \to {1}$. What is more, the DFM can
filter out the statistical fluctuations due to finite number of
eigenvalues in the spectrum of a network.

The adjacency matrices are diagonalized with the Matlab version of
the software package PROPACK [25].

\section{\label{sec:level1}Results}

\begin{figure}
\scalebox{0.8}[0.8]{\includegraphics{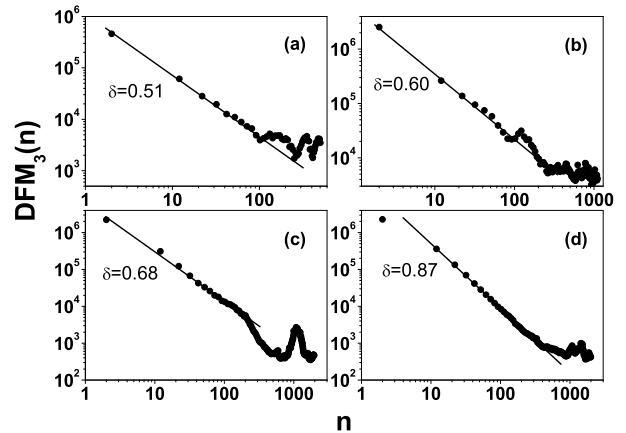}}
\caption{\label{fig:epsart} Four typical results for the
Erdos-Renyi network model.The parameter $q=3$. Denote the size of
a network with $N$.
   (a)$N=10^4$, $p_{ER}=\frac{0.8}{N}<p_{c}$. We have $\delta=0.51$,
   which is consistent with the random behavior of the spectrum.
   The corresponding PDF is Gaussian. (b) $N=10^4$, $p_{ER}=\frac{1}{N}=p_{c}$.
   We have $\delta=0.60$, a slight deviation from random. For (c) and (d)
   $(N,p_{ER},\delta)=(4\times 10^3,\frac{4}{N},0.68)$ and
   $(4\times 10^3,\frac{8}{N},0.87)$, respectively.}
\end{figure}

Consider firstly the Erdos-Renyi model [26,27]. Starting with $N$
nodes and no edges, connect each pair with probability $p_{ER}$.
For $p_{ER} < \raise0.5ex\hbox{$\scriptstyle
1$}\kern-0.1em/\kern-0.15em\lower0.25ex\hbox{$\scriptstyle N$}$
the network is broken into many small clusters, while for $p_{ER}
\ge \raise0.5ex\hbox{$\scriptstyle
1$}\kern-0.1em/\kern-0.15em\lower0.25ex\hbox{$\scriptstyle N$}$ a
large cluster can be formed, which in the asymptotic limit
contains all nodes [27]. $p_c = \raise0.5ex\hbox{$\scriptstyle
1$}\kern-0.1em/\kern-0.15em\lower0.25ex\hbox{$\scriptstyle N$}$ is
a critical point for this kind of random networks.

Fig.1 presents four typical results for Erdos-Renyi networks. For
$p_{ER} < \raise0.5ex\hbox{$\scriptstyle
1$}\kern-0.1em/\kern-0.15em\lower0.25ex\hbox{$\scriptstyle N$}$,
the scaling exponent is,$\delta = 0.51$, which is consistent with
the random behavior of the spectrum. With the increase of $p_{ER}
$, $\delta $ becomes larger and larger. The spectrum tends to
display a significant scale invariance.

  As one of the most widely accepted models to capture the
clustering effects in real world networks, the WS small world
model has been investigated in detail [1,28-31]. Here we adopt the
one-dimensional lattice model. Take a one-dimensional lattice of
$L$ nodes with periodic boundary conditions, and join each node
with its $k$ right-handed nearest neighbors. Going through each
edge in turn and with probability $p_r $ rewiring one end of this
edge to a new node chosen randomly. During the rewiring procedure
double edges and self-edges are forbidden.

\begin{figure}
\scalebox{0.8}[0.8]{\includegraphics{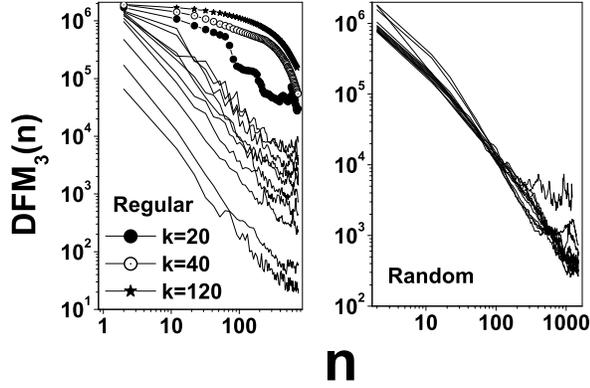}}
\caption{\label{fig:epsart} DFM for the two extreme conditions of
the WS network model, i.e., regular networks ($p_r=0$) and the
corresponding completely random networks ($p_r=1$). The size of a
network $N=3000$. And $q=3$. For these generated networks, when
the number of right-handed neighbors $k$ is small
($k=1,2,\cdot\cdot\cdot,9$) the DFMs obey a power-law.}
\end{figure}

\begin{figure}
\scalebox{0.8}[0.8]{\includegraphics{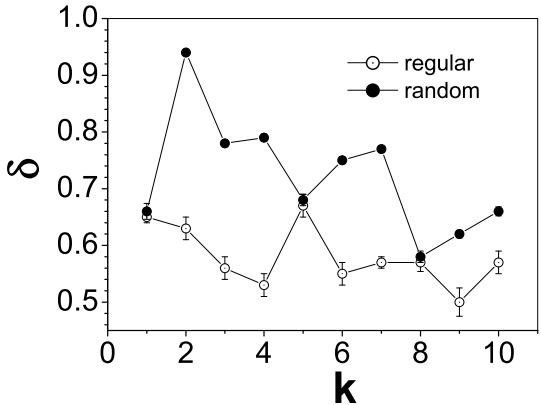}}
\caption{\label{fig:epsart} The values of the exponent $\delta$
for the two extreme conditions of the WS network model, i.e, the
regular networks ($p_r=0$) and the corresponding completely random
networks ($p_r=1$). The size of a network $N=3000$. And $q=3$. The
values of $\delta$ for the regular networks are in the range of
$0.58\pm 0.08$, a slight deviation from $0.5$ that corresponds to
a Gaussian distribution. The values of $\delta$ for the completely
random networks are significantly larger than that of the
corresponding regular networks (with a few exceptions). }
\end{figure}

  Fig.2 and Fig.3 show the results for two extreme conditions of the WS network
 model, i.e.,  the regular networks with different right-handed neighbors
 ($p_r=0$) and the corresponding completely rewired networks ($p_r=1$).
When the value of $k$ is unreasonable large ($k=20,40,80,120$),
the DFM will not obey a power-law. The scaling exponents for the
regular networks are basically in the range of $\delta=0.58\pm
0.08$, a slight deviation from that of the Gaussion distribution.
The scaling exponents for the completely rewired networks with
$k=2,3,4,6,7$ are significantly larger than that of the
corresponding regular networks.

  Four typical results for the networks generated with the WS model
with different rewiring probability values, as shown in Fig.4,
illustrate the significant scale invariance in spectra of these WS
networks.The values of $\delta $ for these generated networks with
$k=2$ and $k=5$ are presented in Fig5 and Fig.6, respectively. We
are specially interested in the rough range of $p_r \in [0.05,2]$
where the WS model can capture the characteristics of real world
networks. For the generated networks with $k=2$, in the range of
$p_r \in [0.05,0.2]$ we have $\delta \in 0.71\pm 0.05$. And in the
condition of $k=5$, $\delta$ is $0.85 \pm 0.05$ in the range of
$p_r \in [0.1,0.2]$.

\begin{figure}
\scalebox{0.8}[0.8]{\includegraphics{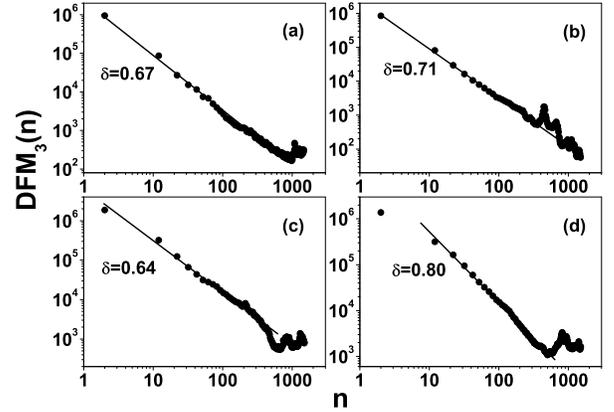}}
\caption{\label{fig:epsart} Four typical results for WS
small-world model. The parameters $q=3$, $k=2$. The size of a
network is $3000$ . (a) $p_r=0.0$, $\delta=0.67$; (b) $p_r=0.1$,
$\delta=0.71$; (c) $p_r=0.3$, $\delta=0.64$; (d)$p_r=0.8$,
$\delta=0.80$.}
\end{figure}

\begin{figure}
\scalebox{0.8}[0.8]{\includegraphics{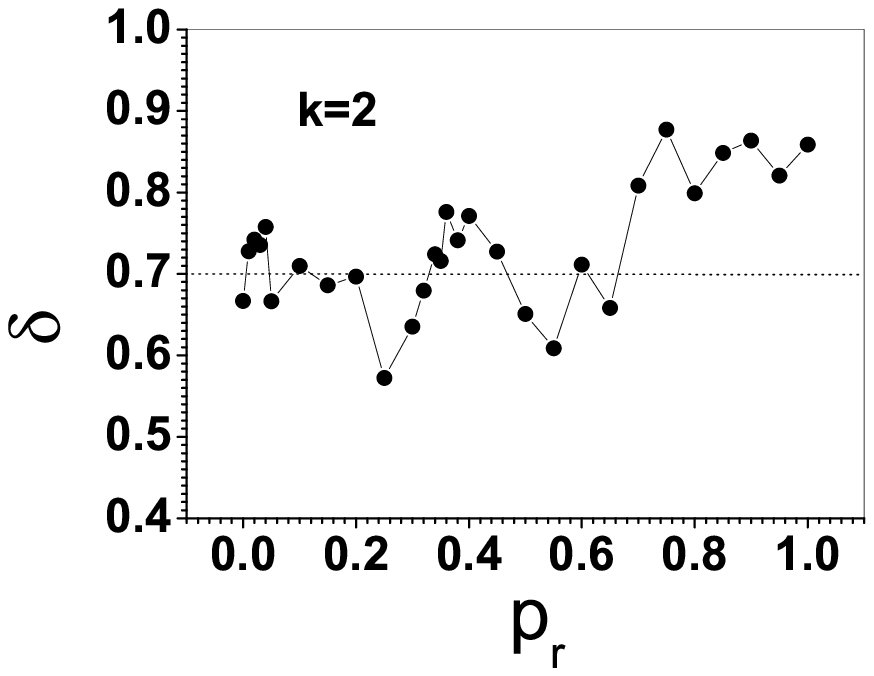}}
\caption{\label{fig:epsart} The values of $\delta$ for generated
WS networks with different rewiring probabilities. The parameters
$q=3, k=2$. In the special range of $p_r \in [0.05,0.2]$, where
the WS small world network model can capture the characteristics
of real world complex networks, we have $\delta \in 0.71\pm
0.05$.}
\end{figure}

\begin{figure}
\scalebox{0.8}[0.8]{\includegraphics{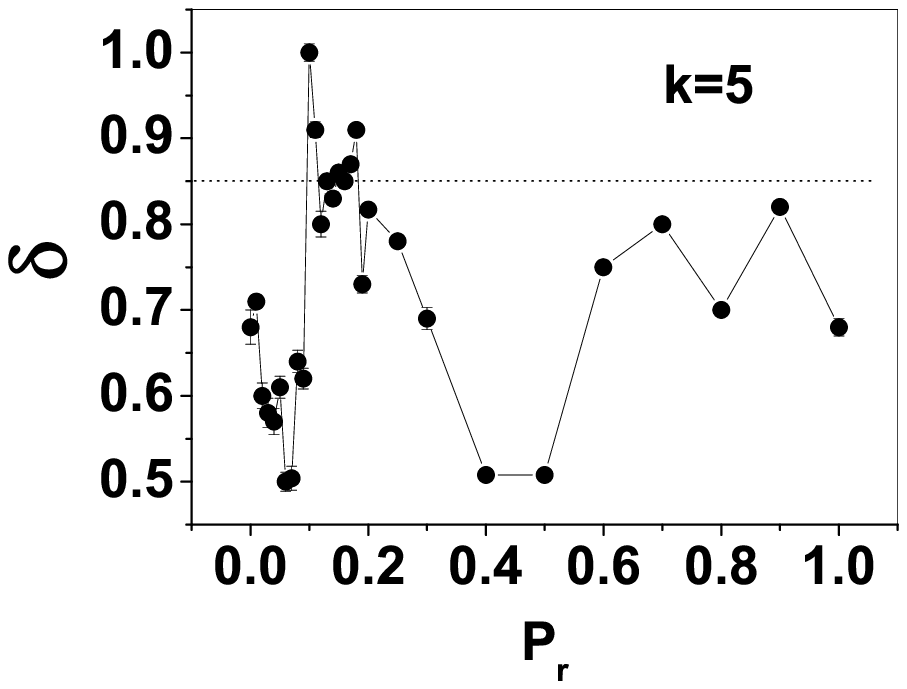}}
\caption{\label{fig:epsart}The values of $\delta$ for generated WS
networks with different rewiring probabilities. The parameters
$q=3, k=5$. In the special range of $p_r \in [0.1,0.2]$, where the
WS small world network model can capture the characteristics of
real world complex networks, we have $\delta \in 0.85\pm 0.05$.}
\end{figure}

\begin{figure}
\scalebox{0.8}[0.8]{\includegraphics{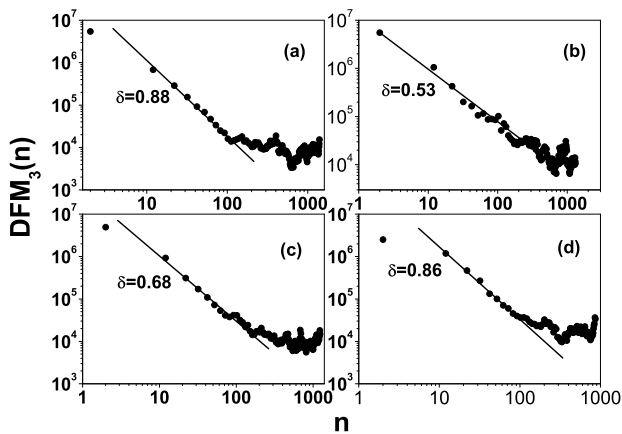}}
\caption{\label{fig:epsart} Four typical results for networks
generated with the GRN model. The parameter $q=3$. The size of a
network is $4000$ . (a) $\theta=0.0$, $\delta=0.88$; (b)
$\theta=0.33$, $\delta=0.53$; (c) $\theta=0.5$, $\delta=0.68$;
(d)$\theta=1.0$, $\delta=0.86$.}
\end{figure}

\begin{figure}
\scalebox{0.8}[0.8]{\includegraphics{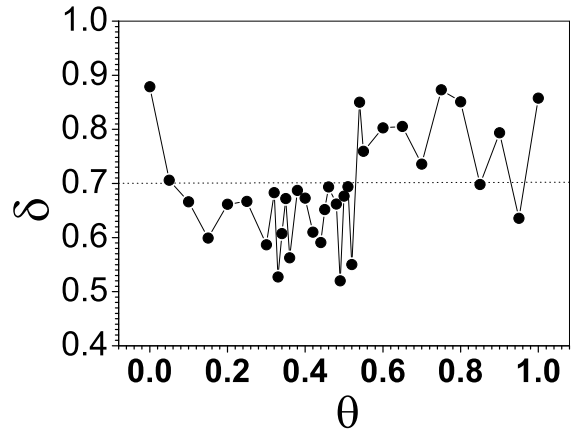}}
\caption{\label{fig:epsart} The values of $\delta$ for the
networks generated with the GRN model with different $\theta$. At
two points $\theta = 0.33,0.49$, we have $\delta = 0.53$ and
$0.52$ (two minimum values), respectively. In the range of $\theta
\in [0.33,0.49]$, we have $\delta = 0.6\pm 0.1$. And the value of
$\delta$ oscillates around $\delta = 0.6$ abruptly. In the range
of $\theta \in [0.54,1]$, we have basically $\delta > 0.7$.}
\end{figure}

  Consider thirdly the growing random network (GRN) model [30,32].
Take several connected nodes as a seed. At each time step, a new
node is added and a link to one of the earlier nodes is created.
The connection kernel $A_k $, defined as the probability that a
newly introduced node links to a pre-existing node with $k$ links,
determines the structure of this network. The considered complex
networks are generated with a class of homogeneous connection
kernels, $A_k \propto k^\theta (0 \le \theta \le 1)$.

The arguments in literature [32] show that there are two critical
points at $\theta _1 = \frac{1}{3}$ and $\theta _2 = \frac{1}{2}$,
which separate the networks into four groups. The four groups are
$\left[ {0,\frac{1}{3}} \right)$,$\left( {\frac{1}{3},\frac{1}{2}}
\right)$,$\frac{1}{2}$and $\left( {\frac{1}{2},1} \right)$. From
the values of $\delta $ for GRN networks with different $\theta $,
shown in Fig.6, we can find that at two points $\theta =
0.33,0.49$, we have $\delta = 0.53$ and $0.52$ (two minimum
values), respectively. In the range of $\theta \in [0.33,0.49]$,
we have $\delta = 0.6\pm 0.1$. And the value of $\delta$
oscillates around $\delta = 0.6$ abruptly. In the range of $\theta
\in [0.54,1]$, we have basically $\delta > 0.7$.

\section{\label{sec:level1}Summary}
In summary, we introduced a new concept called DFM and use it to
reveal scale invariance features embedded in spectra of complex
networks. For an Erdos-Renyi network with connecting probability
$p_{ER} < \frac{1}{N}$, the scaling exponent is $\delta = 0.5$,
while for $p_{ER} \ge \frac{1}{N}$ the scaling exponent deviates
from $0.5$ significantly. For the regular networks generated with
the WS model with $p_r=0$, the scaling exponents deviate slight
from $0.5$, the value corresponding to the Gaussian PDF. The other
extreme condition is that the $\delta$ values for the random
networks generated with the WS model with $p_r=1$ are basically
significant larger than that for the corresponding regular
networks (there are few exceptions). In the specially interested
range of $p_r \in [0.05,0.2]$, where the WS model can capture the
properties of real world networks, the spectra display a typical
scale invariance. Two critical points are found for GRN (growing
random network) networks at $\theta = 0.33$ and $\theta = 0.49$,
at which we have two minimum values of $\delta = 0.53, 0.52$,
respectively. In the range of $\theta \in [0.54,1]$, we have
basically $\delta
> 0.7$. Hence we find self-similar structures in all the spectra
of the considered three complex network models. This common
feature may be used as a new measurement of complex networks in a
unified way. Comparison with the regular networks and the
Erdos-Renyi networks with $p_{ER} <p_c=\frac{1}{N}$ tells us that
this self-similarity is non-trivial.

  The self-similar structures in spectra shed light on the scale symmetries
embedded in the topological structures of complex networks, which
can be used to obtain the possible generating mechanism of complex
networks. Quasicrystal theory tells us that the aperiodic
structure of lattice will induce a fractal structure in the
corresponding spectrum. The most possible candidate feature
sharing by all the complex networks constructed with the three
models may be fractal characteristic, which has been proved in a
very recent paper [33]. Based upon this feature, we may construct
a unified model of complex networks.

\begin{acknowledgments}
 This work is supported by the Innovation Fund of Nankai
 University. One of the authors (H. Yang) would like to thank
 Prof. Yizhong Zhuo, Prof. Jianzhong Gu in China Institute of
 Atomic Energy for stimulating discussions.
\end{acknowledgments}

\end{document}